\newcommand{\CLASSINPUTbottomtextmargin}{0.7in}
\documentclass[journal]{IEEEtran}

\usepackage{cite}

\ifCLASSINFOpdf
  \usepackage[pdftex]{graphicx}
  \DeclareGraphicsExtensions{.pdf,.jpeg,.png}
\else
  \usepackage[dvips]{graphicx}
  \DeclareGraphicsExtensions{.eps}
\fi

\usepackage[fleqn]{amsmath}
\usepackage[utf8]{inputenc}
\usepackage{color}

\usepackage{textcomp}
\usepackage{siunitx}
\usepackage{multirow}
\usepackage{adjustbox}

\usepackage{url}

\date{June 22, 2019}

\newcommand{\add}[1]{{\color{black} #1}}
\usepackage{soul}

\usepackage{tikz}

\newcommand\copyrighttext{%
  \footnotesize \textcopyright~2019 IEEE. Personal use of this material is permitted. Permission from IEEE must be obtained for all other uses, in any current or future media, including reprinting/republishing this material for advertising or promotional purposes, creating new collective works, for resale or redistribution to servers or lists, or reuse of any copyrighted component of this work in other works. DOI: \texttt{10.1109/TNSM.2019.2925506}, IEEE}
\newcommand\copyrightnotice{%
\begin{tikzpicture}[remember picture,overlay]
\node[anchor=south,yshift=10pt] at (current page.south) {\fbox{\parbox{\dimexpr\textwidth-\fboxsep-\fboxrule\relax}{\copyrighttext}}};
\end{tikzpicture}%
}

\begin{document}
%
\title{Fast Decision Algorithms for Efficient Access Point Assignment in SDN-Controlled Wireless Access Networks}
%
%
%

\author{Pablo Fondo-Ferreiro, Saber Mhiri, Cristina López-Bravo, Francisco J. González-Castaño, Felipe Gil-Castiñeira }

%
%

\markboth{DRAFT}%
{DRAFT}
%



\maketitle
\copyrightnotice

\begin{abstract}

Global optimization of access point (AP) assignment to user terminals requires efficient  monitoring of user behavior, fast decision algorithms, efficient control signaling and fast access point reassignment mechanisms. In this scenario, Software Defined Networking (SDN)  technology may be suitable for network monitoring, signaling and control. We recently proposed embedding virtual switches in  user terminals for direct management by an SDN controller, further contributing to SDN-oriented access network optimization. However, since users may restrict terminal-side traffic monitoring for privacy reasons (a common assumption by previous authors), in this work we infer user traffic classes at the access points. On the other hand, since \add{handovers will be more frequent in dense small-cell networks (e.g. mmWave based 5G deployments will require dense network topologies with inter-site distances of $\sim$150--200 meters)}, 
the delay to take assignment decisions should be minimal. 
To this end, we propose taking fast decisions based exclusively on extremely simple network-side application flow-type predictions based on past user behavior. Using real data we show that
a centralized allocation algorithm based on those predictions achieves network utilization levels that approximate those of optimal allocations. We also test a distributed version of this algorithm. Finally, we quantify the elapsed time since a user traffic event takes place until 
its terminal is assigned an access point, when needed.
\end{abstract}


%
\IEEEpeerreviewmaketitle


\section{Introduction}\label{s_motivation} 
\IEEEPARstart{T}he present research is motivated by diverse facts and observations. 


First, new high frequency bands (such as mmWave) will have very short ranges due to blockage \cite{rappaport}. As a result, user terminals will  \add{be more likely to experience handovers in future access networks with high cell densities.  In this scenario, reducing handover time is considered a key strategy to avoid throughput degradation \cite{plosone, horate}. In addition, this scenario will also be troublesome for user applications with stringent latency constraints in case of long handover times, since these may led to unacceptably long worst-case data delays. This is the case, for example, of real-time gaming \cite{vrband} and, in a relatively close future, cloud-rendered virtual reality \cite{vrband,metis}}. 
 Therefore, handover times \add{(which include the decision and the exchange of control messages between the participating network elements \cite{horate,3gpp38300})} will have to be extremely short. 
Second,  for current bandwidth-hungry applications (e.g. 4K video  \cite{metis}, for example), bandwidth will not be limiting in principle as mmWave-enabled 5G wireless networks have a declared goal of gigabit-per-second access bandwidths \cite{rappaport}. However, new applications may be several orders of magnitude more demanding, such as virtual reality with high-quality rendering at the edge \cite{vrband}. Therefore, AP assignment will focus on efficient service provision with an unprecedented focus on  latency requirements while satisfying bandwidth constraints.
Third, \add{in the context of Wi-Fi networks, standards such as  IEEE 802.11r and IEEE 802.11k \cite{ieee80211r, ieee80211k}, and some proprietary solutions (such as Instant by Aruba and Meraki by Cisco) have improved  handover and AP assignment procedures. However, 
they only consider terminal-side handovers, which cannot achieve a global network optimization.} 

This context has lead us to formulating the following question:

{\it Is it possible to take very fast decisions at a centralized controller of a SDN-enabled WLAN to assign user connections to access points while satisfying high network utilization levels?} 

Here, a key issue is what we understand by ``fast''. In general, as commented in the abstract, optimal access point assignments to user terminals require efficient  monitoring of user behavior, fast decision algorithms, efficient control signaling, and fast access point reassignment mechanisms. Software Defined Networking (SDN) technology may be suitable for network monitoring, signaling and control \cite{Sakir,Giraldo}.

\enlargethispage{-1\baselineskip}

However,  since terminal-side information on traffic monitoring may not be available for privacy reasons (a common assumption in the review  in Section \ref{s_related_work}),  network-side decision algorithms must rely on  estimations of user traffic classes at the access points. Accurate traffic classifications are feasible within just seconds after the user initiates transmission \cite{Raschella}, but,
since 
\add{handovers will be more frequent in 5G small-cell networks due to short ranges and cell densification, that classification time 
may be unacceptable. In particular, handover decision time will be critical in 5G mmWave small-cell networks due to the characteristics of mmWave bands, which stress the need of planned or predictive handovers
~\cite{5glatency}. Under these circumstances, a long handover decision  could even imply connectivity loss \cite{Billen} and thus increased worst-case traffic latency, which would be unacceptable for the service level agreements of latency-sensitive applications. Consequently,} in this work we propose taking {\it fast decisions drawing exclusively on extremely simple network-side  flow type predictions based on past  user behavior}. Specifically, the novelty of our proposal is that, instead of analyzing the traffic class of a new flow after it starts, a traffic class prediction is available right when it starts based on the past history of the user. Thus, the time to make an AP/small-cell assignment decision involves the time it takes to consult the traffic class we expect to receive, instead of the time to analyze it in real time, \add{contributing to reduce overall handover time}. Traffic data, however, must be gathered and analyzed in the background, completing the user's history and improving future predictions.

Another interesting aspect of our research is that we evaluate our approach with real data.  
We derived a first scenario from the publicly available LiveLab dataset \cite{LiveLab} collected in a large Wi-Fi campus network with real users. 

However, inspired on that dataset we then propose a second speculative 5G ultra dense network scenario. We show that in both scenarios a fast centralized prediction algorithm based on past history achieves network utilization levels  that approximate those of optimal allocations.

We also test a distributed version of the algorithm. Finally, to show that very fast decisions are feasible, we quantify the elapsed time since a user traffic event takes place until 
its terminal is assigned an access point as a result of an allocation decision, assuming SDN monitoring and control. 

The rest of this paper is organized as follows: in section \ref{s_related_work} we discuss the background of this research. In sections \ref{s_approach}  and \ref{s_evaluation} we describe and evaluate our approach. Section \ref{s_conclusion} concludes the paper.

\section{Related Work}\label{s_related_work}
%
%
%
%
\subsection{Network Context}

The problem of finding an optimal attachment of terminals to wireless APs has been extensively addressed in the literature \cite{Yen,Chen13,Ernst,Antonopoulos}. Most approaches, as stated in \cite{Raschella}, try to assign each terminal to the best AP in terms of a single performance metric, such as RSSI, spare throughput, or bandwidth efficiency.

Especially in the case of IEEE 802.11 networks, it has often been assumed that resources are severely limited, and this may still be the case for the more demanding next generation services in 5G networks \cite{vrband}. Hence, new approaches are needed to meet QoS and QoE requirements of diverse, rich mobile applications in next-generation dense cellular networks. 

There exist different proposals to manage handovers efficienty in 5G ultra-dense networks (UDN) with very short transmission paths \cite{XGe}, including keeping existing connections  before establishing new ones \cite{Park}; setting simultaneous connections via heterogeneous technologies (which a local coordinator will switch  in the event of link failures)\cite{Polese}; and analyzing the probability of a user staying within coverage of the beam for a given time \cite{JChen}. 

Recently, some authors have proposed using SDN to control AP selection dynamically \cite{Raschella,Sood,Lee,Chen17}. SDN has several advantages: (\textit{i}) QoS management is intrinsic to it; (\textit{ii}) SDN-based architectures can manage heterogeneous access networks (WLAN, WiMAX, ZigBee, 4G, 5G);  and (\textit{iii}) an SDN-centralized controller can visualize the status of the whole network or a significant region of it, meaning that SDN-driven AP assignment may contribute to global network optimization.

\subsection{Access Point Assignment}

Two opposing philosophies have dominated AP assignment  so far: network-controlled methods in low-density cellular networks with QoS guarantees, on the one hand, and terminal-initiated methods in high-density AP deployments without QoS guarantees, on the other. However, even though these philosophies currently coexist in offloading scenarios, the line dividing them is blurred. 

In \cite{Billen}, the probabilities of terminal transitions  to neighbor eNBs and their available resources are modeled with a Markov chain. A controller  determines the transition probabilities between states, which represent  available resources at  different eNBs. The next eNB is estimated and assigned virtually to a mobile device before the device moves. When the sojourn time for the eNB expires, the mobile node checks the OpenFlow table and sends  a  handover request to the following eNB.

In 
\cite{Sood}, a local controller in the user device receives AP bandwidth estimates from an SDN controller. Based on this estimation, the user device connects to the less loaded AP provided RSSI level is high enough. 

Logically, since the decisions are taken at the terminal side, 
they are suboptimal. 

In \cite{Chen17}, when a mobile station tries to join the network for the first time, the SDN controller gathers information from the APs receiving the mobile station probe requests. The controller then calculates the optimal AP for the mobile station considering a combination of throughput, packet loss rate, and RSSI level. Finally, the controller commands the most suitable AP to respond to the mobile station with a probe response. For those mobile stations that are already connected to the network, the controller checks continuously if the packet loss rates of the corresponding APs exceed a predefined threshold. If they do, some of the mobile stations are reassigned to new APs. Migration decisions are based on the previously mentioned metrics and on the activities of the mobile stations (the activity of a mobile station is the ratio of its contributed throughput to the total throughput of its AP so far). To compute the activity of the mobile stations the APs must keep track of all their incoming packets.

In \cite{Raschella}, an SDN controller assigns APs to the terminals according to users’ QoS requirements. The approach maximizes the suitability of user flows to APs and minimizes the impact of the match on the remaining active flows. Specifically, the approach considers (\textit{i}) the best bit rates an AP can provide to new requests (closest rates to those requested), (\textit{ii}) the requested bit rates, and (\textit{iii}) the impact on other active flows in the network. The authors assume the availability of a network-side method for classifying user flows into QoS categories (for instance Machine Learning \cite{Nguyen}).  In \cite{Lee} the authors extended the approach to also consider the state of backhaul links. Overall, this approach may achieve an optimal assignment at the expense of attaining real-time full knowledge of the network. 

The disadvantage of nearly-globally-optimal 
assignment methods 
without terminal-side information  like \cite{Raschella} is that they require the APs to inspect user packets and process them to determine the flows' QoS categories they need as input. Otherwise, the next AP has to be chosen based on the amount of  spare resources 
as in \cite{Chen17} (and not according to the required QoS).  
Even though in \cite{Raschella} it is stated that it can take just 1 second for an AP to determine the QoS category of a flow, this may be too long for some next-generation application and does not fulfil 5G requirements.

Indeed, some fast network-side classification solutions assume packet-level inspection and a priori knowledge about transport protocol usage   \cite{Bujlow}. These low-level protocol-dependent features may change in the future and may even be encrypted (considering the growing tendency for traffic encryption). In addition, the time since a terminal enters a network until the user starts generating data is unpredictable. In such cases, the terminal should already be assigned to the most convenient AP according to the knowledge at hand.

In our proposal, we explore the possibility of making reassignment decisions based exclusively on the current state of the APs, the last known quality of radio links, and a  fast assessment of users' needs based on their past history of traffic generation/application usage.
A reassignment decision may be triggered by the user terminals (each time they  launch a new application, for example) or by the network. We remark that decisions are taken based solely on past history. Afterwards, in the background, the network side starts a new estimation 
of ``true'' application QoS based on packet processing --whose result is added to the QoS history--.
Terminal involvement is kept to a minimum and prediction algorithms only work with pre-existing  information each time they are invoked. This results in a relaxed, suboptimal assignment, since the assignments are not based on current 
terminal behavior.

\subsection{Application Prediction}

The problem of predicting the next application from a previous history of executed applications has already been addressed. Tan et al.  proposed an algorithm to predict mobile application usage patterns \cite{Tan}. They conducted experiments on the Nokia MDC dataset with traces of 38 users. Their experiments showed periodic patterns that were strongly dependent on recent user actions. Yan et al.  designed a method to preload applications from contextual information such as user location and temporal access patterns \cite{Yan}. The algorithm in \cite{Baeza}, based on a parallelized Tree Augmented Naive Bayes (TAN) algorithm \cite{Friedman}, organizes the screen content of a terminal by displaying the applications the user is most likely to launch. \add{However, we do not need to determine the exact application that is being used, just to guess the application flow type} and thus we can afford much simpler prediction algorithms. This is why we employ our simple user profiler in \cite{Mhiri} based on Markov chains (See section \ref{ss_flowPredictor}), which provides an $F$ score between 0.7 and 1 for the different applications. 

\subsection{Direct Control of User Terminals}

Management approaches for high-density access networks have not yet considered the possibility of controlling terminals directly as SDN nodes. In the Traffic Steering Architecture (TSA) we proposed  in  
\cite{Giraldo}, the terminals become actual SDN nodes. This facilitates the exchange of control commands within the same softwarized architecture, both for monitoring (terminal $\rightarrow$ network) and assignment (network $\rightarrow$ terminal) purposes. By managing terminal connections directly from a centralized entity we can take into account the state of the network when commanding the terminal to use the most adequate AP, reducing handover time and complexity. A clear advantage of choosing  SDN is that it can handle multiple access technologies at the terminal side.
Obviously this requires the user to install an SDN agent in the terminal. It could be argued that this is not realistic,  but our results also hold  for any other fast  network-driven roaming approaches.  Network-driven hard handovers are present in existing cellular communications technologies such as LTE \cite{LTEroam} and, in fact, terminal-side agents are also considered in the work by other researchers \cite{Sood}. 

Regarding scalability in dense environments with many users connected to APs that need to communicate with the controller, note that event-driven signaling only takes place when a user sets an initial connection with the network or launches a new application. Afterwards,  signaling periods should be adjusted according to user mobility in a given scenario. Similar signaling requirements are present in the work  by other researchers  \cite{Raschella, Sood, Chen17} and will be necessary for ultra-dense 5G deployments with short transmission paths, high directivity and blockage  by most surrounding objects including the user's body \cite{Billen}.

\section{Proposed Approach}\label{s_approach}

\subsection{System Architecture}\label{ss_system_architecture}

\begin{figure}[ht!]
\centering
\includegraphics[width=0.5\textwidth]{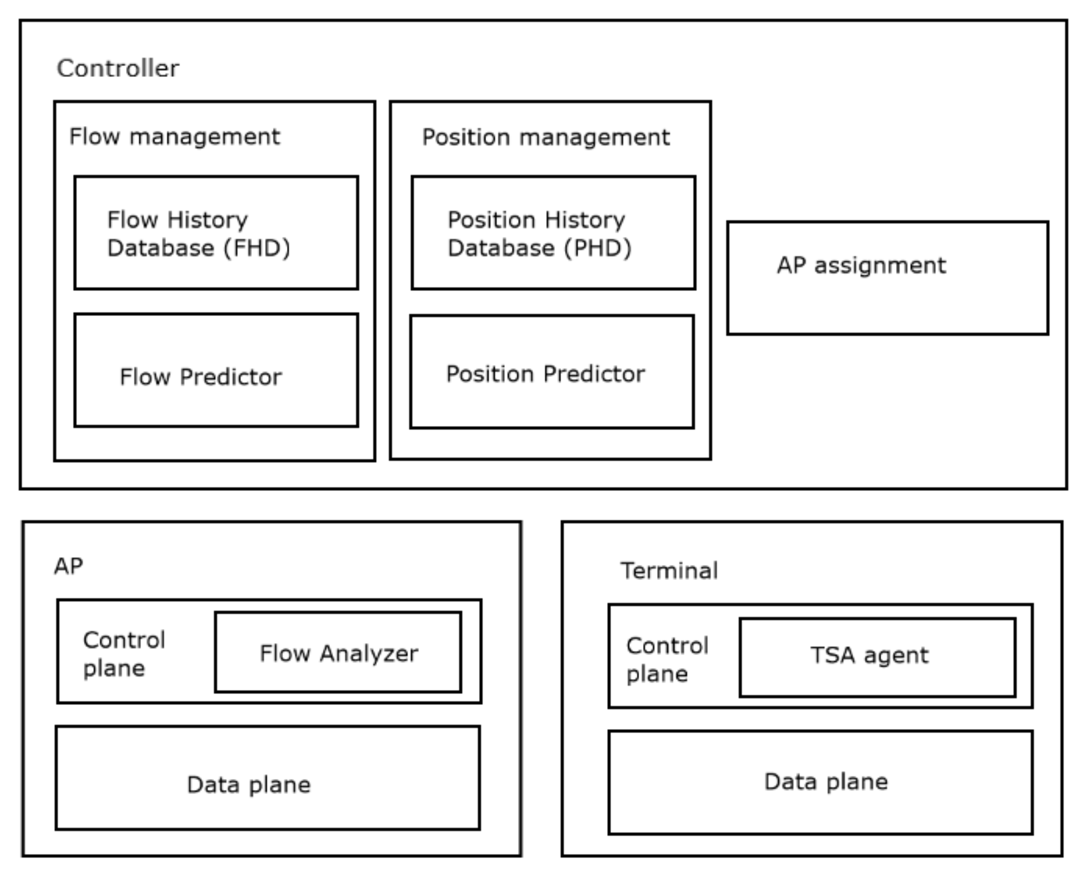}
\caption{System architecture}
\label{Saber1}
\end{figure}

Fig. \ref{Saber1} shows the system architecture.
It comprises the following elements:

\begin{itemize}
\item \textit{Controller}

\begin{itemize}
\item Flow Management module. This module keeps the Flow History Database (FHD) containing past types of terminal flows and their start and end times. With this information, the Flow Predictor sub-module actualizes its transition probabilities to predict the next flow type  whenever a new decision must be made for the corresponding terminal. The FHD also keeps the average application flow rates.
\item Position Management module. This keeps a Position History Database (PHD) containing the 
link qualities between the APs and the terminals as reported by the latter. It may also contain the positions of the terminals if they report these positions periodically. Alternatively, the Position Predictor sub-module predicts the position of each terminal from the last information available in the PHD whenever a new decision must be made.  
\item AP Assignment module. This module executes the centralized AP assignment optimization algorithm, which takes the outputs of the Flow Predictor and the Position Predictor (if available), the average application flow rates and  the last known quality of the links  as inputs to decide the next assignments of access points to terminals.

\end{itemize}

\item\textit{AP}

\begin{itemize}
\item Flow Analyzer module. This module analyzes terminal flows in the background to characterize them. The information is then sent to the controller to be stored in the FHD.
\end{itemize}

\item \textit{Terminal}

\begin{itemize}
\item TSA Agent. This agent sets connections with the APs as commanded by the controller. It may also send terminal information to the controller, if allowed by the user to do so. In  such a case, flow analysis (by the flow Analyzer module) or position prediction (by the position Management module) may be unnecessary.

\end{itemize}
\end{itemize}

\begin{figure}[ht!]
\centering
\includegraphics[width=0.5\textwidth]{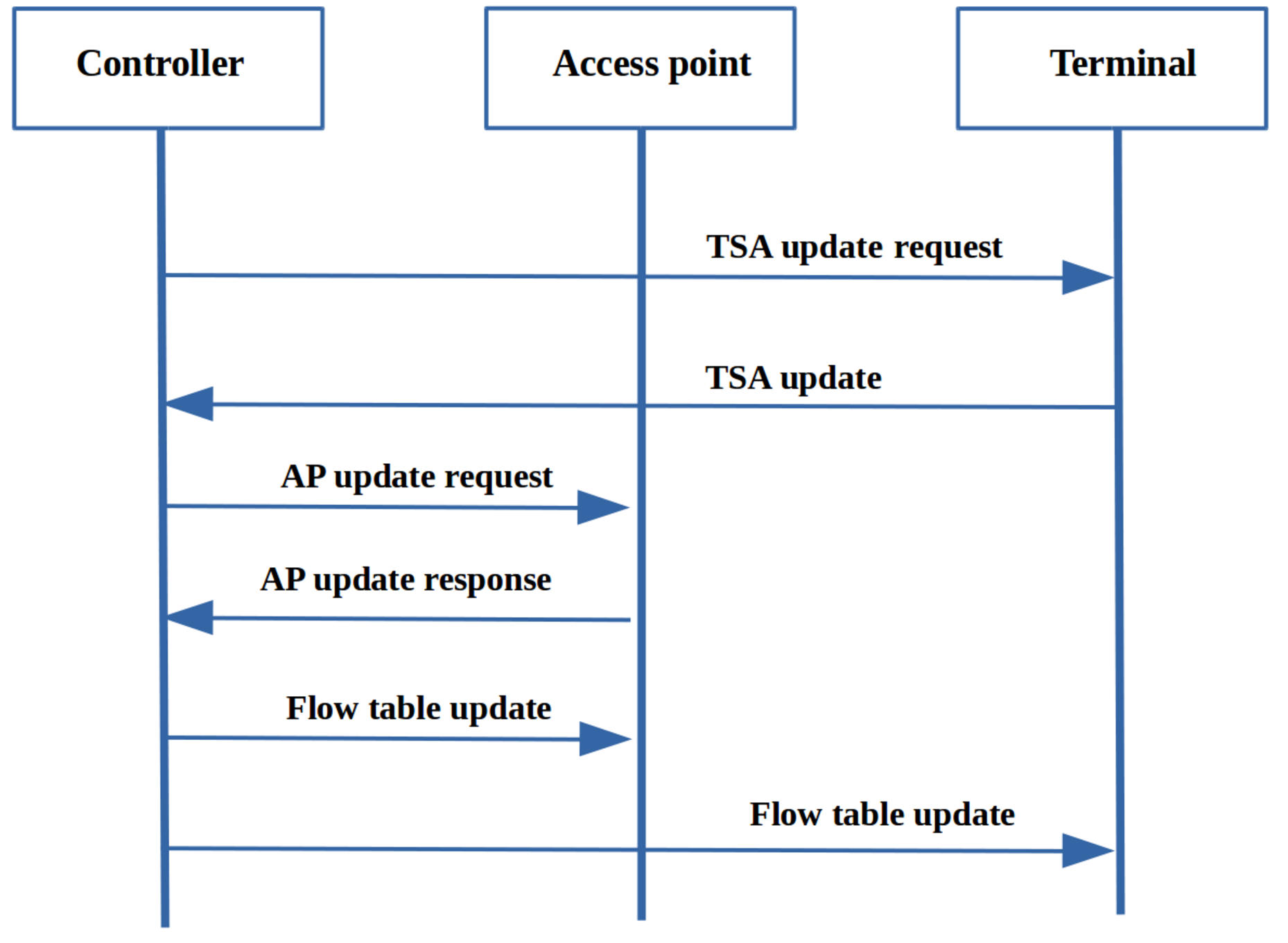}
\caption{Periodic TSA signaling}
\label{Saber2}
\end{figure}

\begin{figure}[ht!]
\centering
\includegraphics[width=0.5\textwidth]{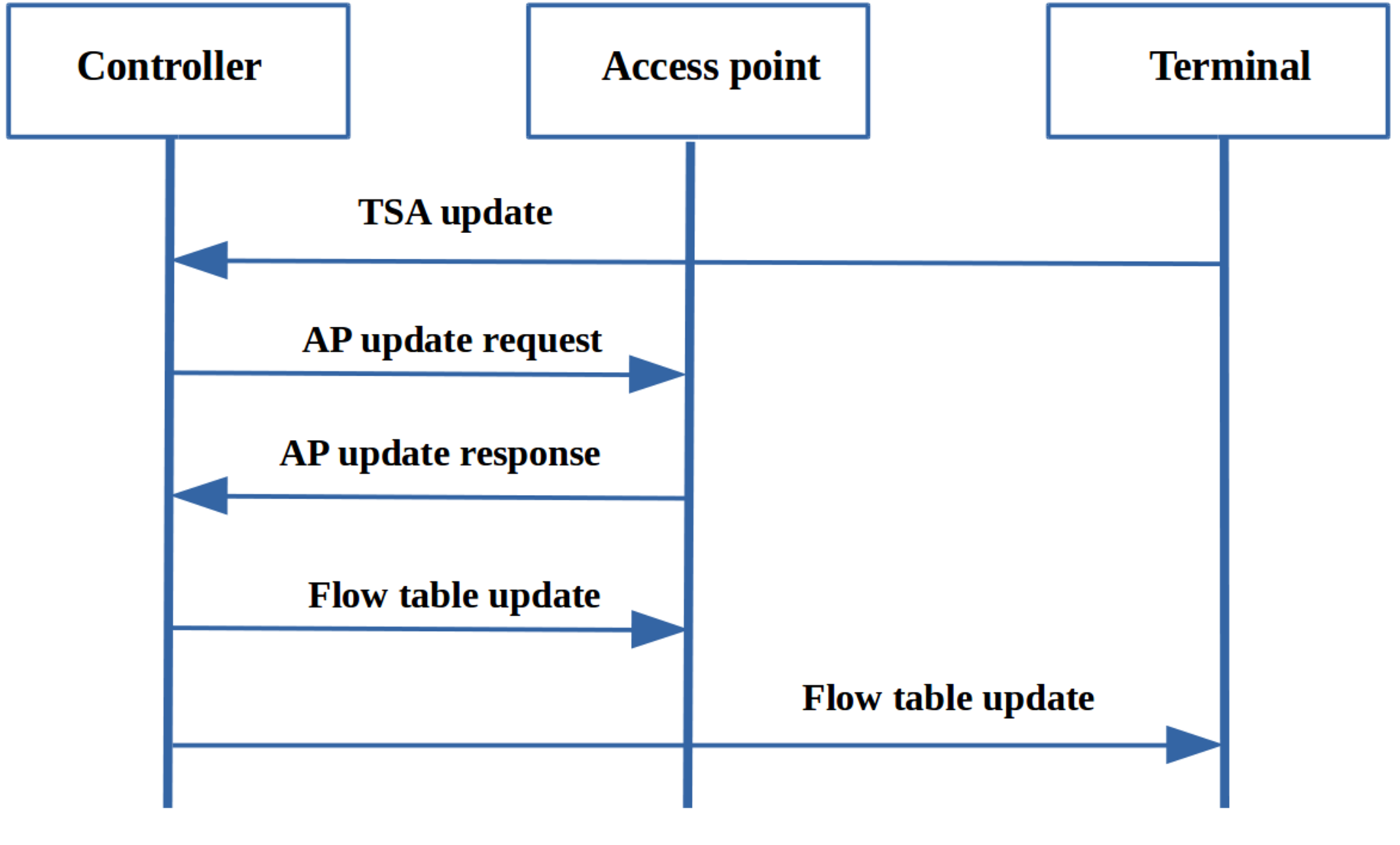}
\caption{Event-driven TSA signaling}
\label{Saber3}
\end{figure}

The TSA agents in the terminals may perform periodic (Fig. \ref{Saber2}) and/or event-driven (Fig. \ref{Saber3}) signaling. The former keeps the FHD and PHD databases of the controller updated on terminal activities and link qualities. Specifically, the controller sends {\it AP update requests} to the APs to check their spare bandwidth and the outcome of their flow analyses. The APs return {\it AP update responses} with their spare bandwidth and the  estimated flow types,  which they tag with the L2 identifiers of the corresponding terminals. The controller also sends {\it TSA update requests}  to the terminals if the user allows them to report their activity. In that case, the terminals return positions, link qualities and/or applications (therefore, flows) types in {\it TSA updates}.

The controller sends SDN {\it flow table updates} to the APs and the terminals to assign AP connections.

Event-driven signaling  allows the controller to react to sudden changes in terminal activities, ensuring  real-time FHD updating. In case of an event (as soon as the terminal requests a connection to the network or the user launches a new application), the TSA agent in the terminal sends a {\it TSA update} with data such as bandwidth consumption, application type, GPS position and accelerometer measurements, if authorized by the user to do. In response to an event, the controller sends {\it AP update requests} to the APs.

As in the case of periodic TSA signaling, the APs report to the controller their spare bandwidth and the estimated flow types in {\it AP update responses}. In the background, they keep estimating the flow types of active connections. These estimations are sent to the controller in subsequent periodic TSA signaling. Once the controller makes a new decision, it updates the terminal and AP flow tables with {\it flow table updates}.

\subsection{Implementation}

Our approach can be implemented with any wireless technology. Nevertheless, the SDN paradigm provides most of the elements required, including a centralized controller (which receives information and handles the different SDN-enabled devices in the network).

The SDN controller is implemented as a network application. Its NorthBound API \cite{sdnreview} (the interface with the applications and services running over the network) receives information on the traffic flows generated by the terminals and allows  the controller to assign an AP to a terminal.  We thus take advantage of existing flow monitoring by SDN-enabled switches and routers. SDN controllers also manage the routing tables of these different network elements. This feature can be used to assign a particular AP to a terminal, and also to guarantee terminal session continuity even when traffic routes are altered (due to changes in serving APs or other route modifications).

Our architecture should be implemented with SDN-capable APs configured to identify new traffic flows and notify the controller accordingly. They can be constructed on APs with enough computing power and OpenWRT support, by  installing OpenvSwitches in them. Note that these are open source tools. We have created such SDN-capable APs on TP-LINK Archer C2600 routers. Furthermore, some commercial SDN-capable wireless access points are available, such as Zodiac WX from Northbound Networks\footnote{https://northboundnetworks.com/products/zodiac-wx}. 

In \cite{Giraldo} we describe and evaluate  a virtual SDN-enabled switch that is embedded in a terminal.  With this approach it is possible to command a terminal with \add{SDN protocols such as Openflow \cite{DTNWifi}} to select a Wi-Fi AP or any other network interface to route its traffic. The internal switch also sends packets to the  controller  when  they  do  not  match  any of the rules  in  the  flow  table, or when they match a table entry with an ``output to controller'' rule. The terminal can also transmit other relevant information such as  location  and link quality levels to the controller using any other protocol (for example, a REST call). In brief, the terminal TSA agent is just an SDN router managing the different communication interfaces of a smartphone, for example.

\subsection{Flow Predictor}
\label{ss_flowPredictor}

The flow predictor of the flow management module employs a simple Markov chain  to guess  the application flow a  terminal is likely to produce next \cite{Mhiri}. The weight between two flow-type nodes is incremented in one unit whenever the application flow represented by the destination node becomes active after the application flow represented by the origin node. Fig. \ref{Saber4} shows an example.  We refer the reader to \cite{Mhiri} for a detailed description of the prediction algorithm and its training. We remark that the approach is valid whether the APs estimate application traffic classes with a flow analyzer or whether the terminals report these directly (in this second case the controller might predict  the next application type {\it before} the terminal reports its use). Application flow types are taken, for example, from a set of representative applications (the example in Fig. \ref{Saber4} has four types). In the scenarios in Section \ref{simresults}, before calculating the chain weights, we applied random under- and oversampling to  historical data \cite{Kotsiantis}. In practice, this could be performed periodically.

\setcounter{figure}{3} 
\begin{figure}[ht!]
\centering
\includegraphics[width=0.5\textwidth]{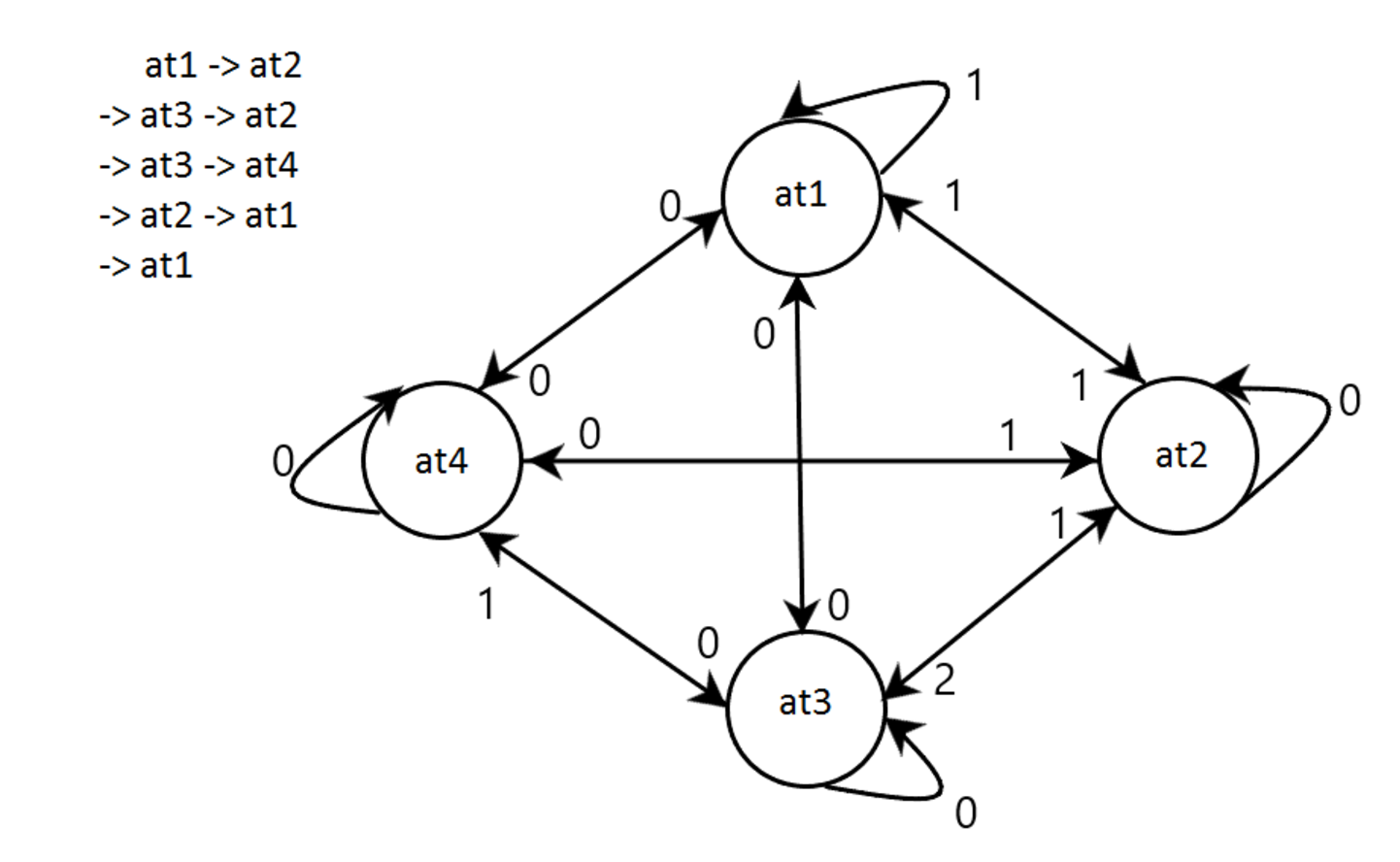}
\caption{Flow predictor. Example with four application flow types}
\label{Saber4}
\end{figure}  

We estimate application flow types and determine application flow rates separately. The former correspond to possible applications flows while the latter correspond to true application rates, which may vary in time (e.g. a user may watch videos with different qualities). Therefore, either the access points or, optionally, the terminals themselves, must monitor average application flow rates  and report them periodically. Whenever the AP assignment module makes a decision, it will employ the last information available.

\subsection{Fitness Function}

 The set of APs is divided in two groups: ``mouse'' APs and ``elephant'' APs. ``Mouse'' APs  do not allow terminals to exchange high-rate traffic for some reason, not necessarily due to limitation of resources. For example, in a campus network,  downloading entertainment streams in laboratory facilities may be forbidden. In a commercial network, access to certain APs may be restricted to users of premium services. 
 
 Let $n$ and $l$ be the respective sizes of the terminal and AP populations in the access network, respectively. Let $E_{AP}$ and $M_{AP}$ be two disjoint sets of indices in \{$1, \cdots,l$\}  (i.e. $E_{AP}$ $\cup$ $M_{AP}$ = \{$1,\cdots,l$\}, $\mid E_{AP}\mid+\mid M_{AP}\mid$ = $l$), such that an AP $j$ only admits flows of type $e$ -elephant- (alternatively $m$ -mouse-)  if $j$ $\in$ $E_{AP}$ (alternatively if  $j$ $\in$ $M_{AP}$). Only exceptionally, if there are no ``mouse'' APs at hand (e.g. those whose indices belong to $M_{AP}$), an ``elephant'' AP in $E_{AP}$ may admit type $m$ flows.   
Similarly, let $E_t[k]$ and $M_t[k]$ be two disjoint sets of indices in \{$1,\cdots,n$\} at event $k$ (i.e. $E_t[k] \cup M_t[k]=\{1,...,n\}$, $\mid E_t[k] \mid + \mid M_t[k] \mid=n$), such that $i \in E_t[k]$ (alternatively $i \in M_t[k]$) if at event $k$ it is predicted that terminal $i$ will generate a flow of type $e$ (alternatively $m$).

We define the following fitness function to be maximized both periodically and when a terminal-side event is triggered:

\begin{equation}
f(\cdot) =  \sum_i \sum_j  L_{ij}[k] q_{ij}[k] r_{ij}[k] 
\label{eq:fittness_funtion}
\end{equation}

Where $L_{ij}[k]=1$ if a connection is set between terminal $i$ and AP $j$ and 0 otherwise, so that  $\sum_i \sum_j L_{ij}[k]$ is the number of admitted flows through the network at event $k$; $q_{ij}[k]$ is a measurement of the quality of the channel between $i$ and $j$ at event $k$;

and $r_{ij}[k]$  is the estimated rate for the predicted flow type from terminal $i$ to AP $j$ at event $k$.

\subsection{Optimization Constraints}

Let us first formulate the  {\it centralized} version of the problem. Let $L[k]$ be a $n\times l$ binary matrix containing the connection proposals at event $k$.  As previously mentioned, $L_{ij}[k]=1$ if a connection between terminal $i$ and AP $j$ is proposed. Matrix $L$ is feasible at event $k$ if:

\begin{equation}
 \forall i,j \mid L_{ij}[k]=1,  q_{ij} > \tau
\label{eq:4}
\end{equation}

\begin{equation}
\sum_j L_{ij}[k] \leq 1 ,\; i=1 \dots n 			
\label{eq:5}
\end{equation}

\begin{equation}
L_{ij}[k]=0 \;\; \textup{if} \;\; (i \in E_t[k] \; \textup{and} \; j \in M_{AP})  
\label{eq:6}
\end{equation}

\begin{equation}
\sum_i L_{ij}[k]r_{ij}[k] \leq R_j[k], j=1\dots l 
\label{eq:7}
\end{equation}

Where $q_{ij}$ is the quality of  the link connecting $i$ with $j$ if $L_{ij}[k]=1$, $\tau$ is a strength threshold and $R_j[k]$ is the room for new flows at AP $j$ at event $k$. 

We only allow  connections between ``mouse'' terminals and ``elephant'' APs when there is no room available for the former in ``mouse'' APs. 

Let $\Omega^m_i[k]$ be the set of ``mouse''  APs in the surroundings of terminal $i$ at event $k$, that is, $j \in \Omega^m_i[k]$ if and only if $j \in M_{AP}$ and $q_{ij}[k] > \tau$.  Then, $\forall i \in M_t[k]$, $ o \in E_{AP}$:

\begin{equation}
L_{io}\sum_{j \in \Omega^m_i[k]} max(R_j[k]-r_{ij}[k],0) = 0 
\label{eq:8}
\end{equation}

In order to improve scalability, it  is possible to formulated a {\it distributed} version of the problem by splitting  (\ref{eq:fittness_funtion})-(\ref{eq:8}) into subproblems corresponding to AP clusters with a different controller each. In case those clusters are loosely connected, many intra-cluster decisions will be globally optimal due to constraint (\ref{eq:4}). The distributed decision will be more relaxed, since a cluster controller  will ignore any APs belonging to other clusters regardless of their proximity to the inter-cluster border, even if they are lightly loaded at event $k$. Let us assume that an efficient clustering into $C$ clusters exists, such that the $c$-th cluster contains the APs in $E_{AP}^c \cup  M_{AP}^c$, $\cup_{c=1\dots C} E_{AP}^c = E_{AP}$, $\cup_{c=1\dots C} M_{AP}^c = M_{AP}$. Then, the $c$-th controller solves the problem obtained by  replacing $E_{AP}$ with $E_{AP}^c$ and $M_{AP}$ with $M_{AP}^c$ in (\ref{eq:fittness_funtion})-(\ref{eq:8}).

\subsection{AP Assignment Algorithm}
\label{ss_AP_Assignment module}

It is straightforward to find points in the feasible region by assigning the terminals  to compatible APs within range (or leaving these terminals unconnected if there are no such APs at hand) as far as (\ref{eq:7}) holds. From a feasibility perspective, a search direction may be generated by picking a terminal and setting an alternative feasible assignment for it.  Several picks may be evaluated in parallel at any iteration. A pick will be accepted if it leads to an improvement in (\ref{eq:fittness_funtion}).

In real networks, active users will seldom move between events that are close in time, so a solution for any given event will quickly allow determining a good feasible point  for the next problem. Also, both for current and foreseen wireless technologies,  (\textit{i}) a terminal will ``see'' a small number of APs within range at any moment, so the dimension of the search subspace will be dominated by $n$, and (\textit{ii}) for many terminals (\ref{eq:7}) will hold no matter what their assignments to surrounding compatible APs are (and therefore, their assignments will be irrelevant for problem (\ref{eq:fittness_funtion})-(\ref{eq:8})). Summing up, as our numerical results confirm, the solution to problem (\ref{eq:fittness_funtion})-(\ref{eq:8}) can be approximated in a  real network in few trials, many of which can be executed in parallel. 

 At event $k$  the search algorithm for the AP assignment module can be formalized as follows, where step 2 could be executed in parallel for each different pick (we omit the event index for the sake of clarity):

\ \\

\noindent {\bf Parallel search optimization algorithm}

\noindent Let $L$ be a feasible initial assignment. Solution $L^*$ is initially set to {\bf 0}.

\noindent {\bf Do in parallel:}

\begin{itemize}
\item[] {\bf Loop} 
\begin{enumerate}
\item Pick $i \in \{1,\dots,n\}$ at random
\item Do:
\begin{itemize}
\item[a)] $\forall k\neq i, \; \forall j$, $L_{kj}'=L_{kj} $; \ \\ $\forall j$ $L_{ij}'=0$
\item[b)] Pick $j \in \Omega_i$ such that $L_{ij}'=1$ is feasible
\end{itemize}
\item If $f(L')>f(L)$ $L=L'$
\item {\bf Coordination step:} Coordinate all processors to share the best assignment $L^*$ so far. If $\mid f(L) - f(L^*) \mid < \epsilon$ {\bf return} $L^*$ {\bf and terminate}. Otherwise, $L=L^*$.

\end{enumerate}

\end{itemize}

\noindent where an AP $j$ belongs to the neighborhood $\Omega_i$ of terminal $i$ if $q_{ij} > \tau$. Parameter $\epsilon$ may be chosen as a stop criterion after which no practical improvement is expected. Logically,  step 4 may be configured to be activated  only after some iterations to reduce coordination load.

\section{Evaluation and Results}
\label{s_evaluation}

\subsection{Evaluation Scenario}
\label{s_scenario}

\subsubsection{Pre-5G LiveLab scenario}

To evaluate our proposal we chose the LiveLab dataset \cite{LiveLab}. This dataset contains  real  data from mobile terminals monitored in the Rice University Wi-Fi campus network.  Dataset features include  the applications running in the terminals, accelerometer measurements and the set of available Wi-Fi APs to each terminal at a given time. At the time of evaluation the campus network had  834 APs covering an area of $700\times500$ $m^{2}$.

\begin{figure}[ht!]
\centering
\includegraphics[width=0.55\textwidth]{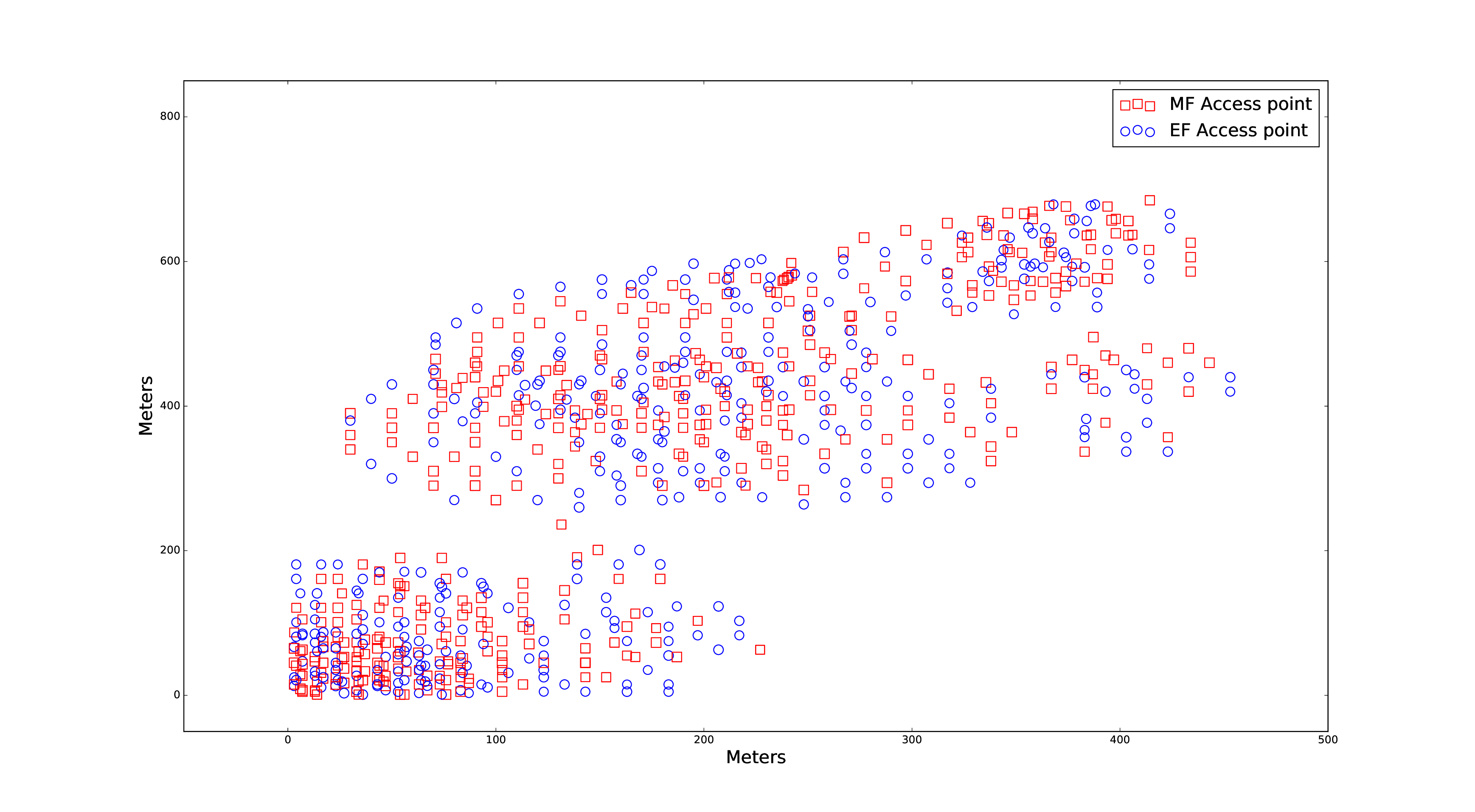}
\caption{Elephant (EF) and mouse (MF) APs in the network.}
\label{Saber5}
\end{figure}

We estimated the topology of the network (shown in Fig. \ref{Saber5}) from these data using the plug \& play approach in \cite{Mhiri}, which allows SDN-controllers to obtain an automatic view of AP locations automatically. The coverage radius of the APs was assumed to be 20 meters indoors.
 
To represent a heavily congested scenario, 60\% of the APs in the network were randomly labeled as mouse APs with spare capacities of about 50 Kbps (that is, capacity left after serving existing terminal connections).

The remaining APs were labeled as elephant APs with spare capacities of 10 Mbps.

\begin{figure}[ht!]
\centering
\includegraphics[width=0.55\textwidth]{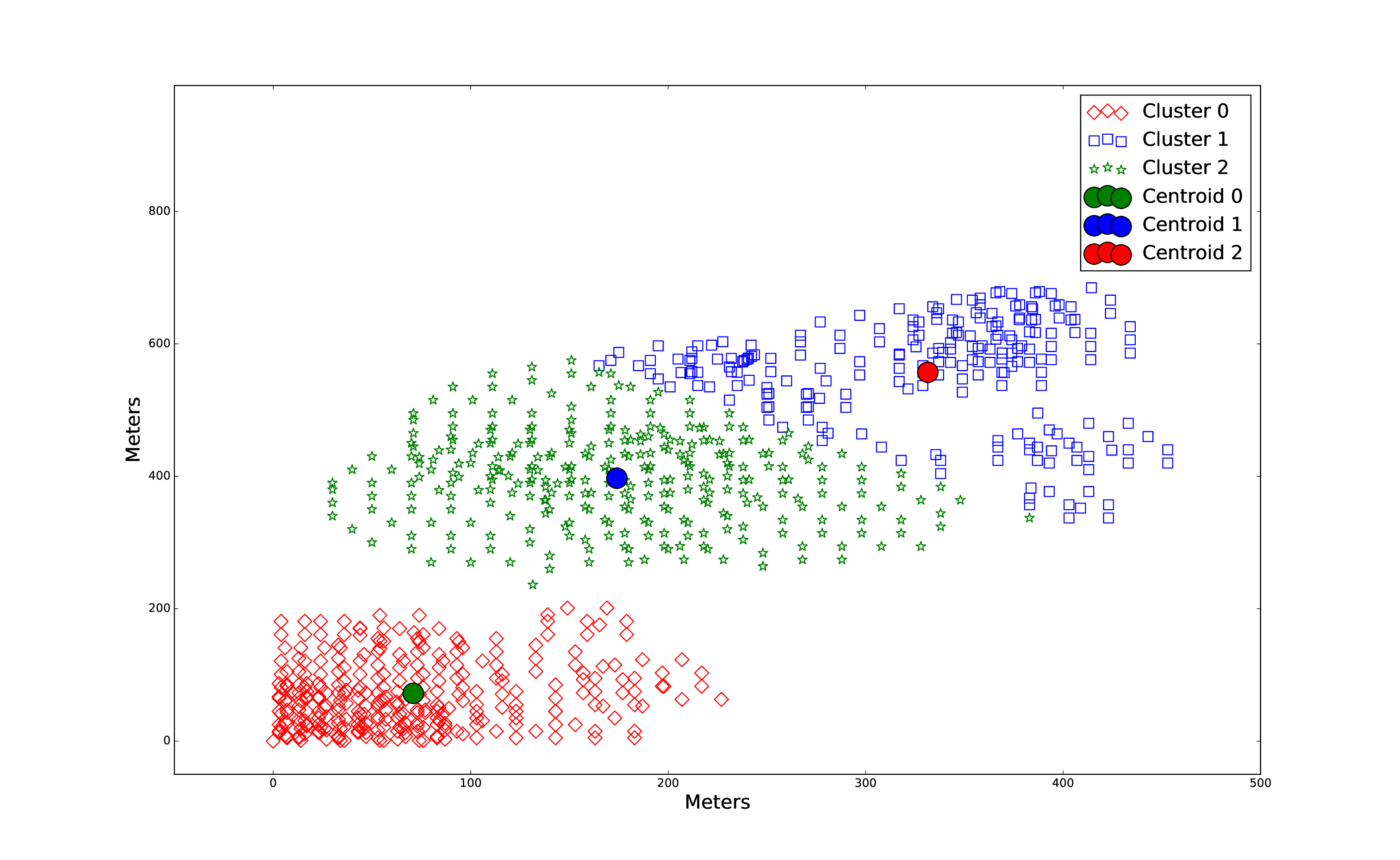}
\caption{Spatial AP clustering for the distributed algorithm}
\label{Saber6}
\end{figure}

To implement the distributed approach we grouped the APs in clusters with low mutual interference. We first guessed  a convenient number of such clusters with the Silhouette Index \cite{Rousseuw},  by successively applying the K-means algorithm \cite{Lloyd} to AP spatial coordinates. Three clusters seemed a good choice. Fig. \ref{Saber6} shows the K-means clustering for this case. Clusters 0, 1 and 2 have 304, 223 and 307 APs, respectively.        

We picked a LiveLab subset comprising the data between 8:00 a.m. and 13:00 p.m of 34 workdays of the Rice University calendar, such that there were data for at least 90 active users each day. In case in a given day there were less than 90 active users we replaced individuals  by groups of users with the same behaviour and mobility.

Even though the LiveLab dataset includes the applications that the terminals run,
it does not include information about application rates. To emulate these  we followed the methodology  in \cite{Mhiri}. First, we discarded any applications that did not have any impact on network occupancy (such as ``gallery'', ``clock'' or ``calendar''). Second, we classified the remaining applications into broader classes: video streaming, calls, messaging, e-mail, sharing and synchronization, entertainment and social applications. Whenever the user launched a new application, we emulated the corresponding  class rate 

 from the traces in \cite{Sengupta}. Finally, we tagged each broader class as ``mouse'' or ``elephant'' depending on the average demanded bandwidth. Table \ref{table_bandwidth} shows the classes  together with their average rates and  their tags \cite{Sengupta}.

 \begin{table}[htb]
\centering
\caption{Application categories, LiveLab dataset}
\label{table_bandwidth}
\begin{adjustbox}{max width=\textwidth}
\begin{tabular}{ | p{3.5cm} | c | c |}
\hline
Application Class & Average Rate  & Application Tag \\
\hline
Video streaming (YouTube, YouTubeLive, Ted, Hostar)  & \multirow {2}{*} {2.58 Mbps}  & \multirow {2}{*} {Elephant} \\ 
\hline
Social media with audio \& video (Hangouts) & \multirow {2}{*} {44.79 Kbps} & \multirow {10}{*} {Mouse} \\
\cline{1-2}
News (TimesNow) & 43.45 Kbps & \\
\cline{1-2}
Sports (StarSport) & 17.73 Kbps & \\
\cline{1-2}
VoIP (Skype, oovoo) & 16.07 Kbps & \\
\cline{1-2}
Social media without audio or video (Facebook, WhatsApp) & \multirow {2}{*} {12.58 Kbps} & \\
\cline{1-2}
Email (Gmail) &  12.58 Kbps & \\
\cline{1-2}
Sharing and synchronization (Dropbox) & \multirow {2}{*} {12.58 Kbps} & \\
\hline
\end{tabular}
\end{adjustbox}
\end{table}
 
 In this case we set $q_{ij}[k] = -\tfrac{1}{\text{RSSI}_{ij}[k]}$, where $\text{RSSI}_{ij}[k] < 0$ is the RSSI between terminal $i$ and AP $j$ at event $k$.
 
 \subsubsection{Speculative 5G scenario}
 \label{spec5Gsce}
 
 The main motivation for evaluating our methodology on a pre-5G \add{ ultra dense network} scenario has been the availability of real user data, unlike for 5G scenarios. However, we consider that our numerical results can be extrapolated to the latter for the following reasons:
 \begin{itemize}
     \item As described in \cite{vrband} (Fig. 5), most applications in 5G networks will be currently existing ones. The main difference is that some existing applications that struggle to achieve their latency requirements with 4G networks, such as real-time gaming and medium-quality augmented reality, will fit better into 5G networks. Indeed, the bandwidth requirements of those applications can be satisfied with current networks. 

     \item The main building blocks of 5G networks will be 5G radio and intelligent network softwarization. The latter may be applied to pre-5G radio access networks (RAN) \cite{5Gchung}.

     \item  The requirements of some applications such as mobile virtual reality and tactile internet, which will still take years to be commercially available, are completely beyond the capabilities of pre-5G networks. However, one of those applications will take a significant share of the expected capacity of a 5G radio small cell \cite{vrband,IMT2020}, that is, a comparable percentage of resources per access point as the “elephant applications” in Table I. They will be devoted to entertainment \cite{pajamental} with similar usage patterns as current video streaming. Finally, they will be served with higher priority by certain access points to guarantee quality of experience requisites.

     \item Dense 5G deployments will require frequent network-controlled handovers \cite{Billen}. The high directivity of 5G radio  implies that interference will be very low and that  quasi-wired models will be valid \cite{pwire1,pwire2,pwire3}, further justifying the interest of the centralized decision algorithms that will be necessary to handle handovers. These decision algorithms will have to satisfy 5G \add{handover} latency constraints.
     
 \end{itemize}

Therefore, we modeled a speculative 5G scenario with the application categories in Table \ref{table_bandwidth_5g}. In it, inspired by \cite{vrband}, we assume that the applications with usage patterns similar to current entertainment video will reach  much higher rates in the order of Gbps (such as virtual reality gaming), and that those with usage patterns similar to current user messaging will reach rates in the order of Mbps due to multimedia attachments or live video. That is, we simply replace the applications of a given category (elephant or mouse) in the filtered LiveLab dataset  by those of the same category in Table \ref{table_bandwidth_5g}, for the same users. The APs in the second scenario keep their types and ranges and can be grouped in the same clusters, but we now set the spare capacities of elephant and mouse APs to 10 Gbps and 10 Mbps, respectively. Logically, the RSSI measures in the LiveLab dataset are no longer valid. By considering the linear macroscopic pathloss estimate (\ref{eq:pathloss}) between terminal $i$ and AP $j$ in \cite{pathloss5G}, where $d(\cdot)$ is the euclidean distance, 

\begin{equation}
g_{i\rightarrow j} = 75.85+37.3 \log_{10} (d(i,j))
\label{eq:pathloss}
\end{equation}

we set $q_{ij}[k]=-\frac{1}{g_{i\rightarrow j}}$.

\begin{table}[htb]
\centering
\caption{Application categories, speculative 5G scenario}
\label{table_bandwidth_5g}
\begin{adjustbox}{max width=\textwidth}
\begin{tabular}{ | p{3.5cm} | c | c |}
\hline
Application Class & Average Rate  & Application Tag \\
\hline
Virtual Reality (VR) entertainment  & \multirow {2}{*} {2.5 Gbps}  & \multirow {2}{*} {Elephant} \\ 
\hline
High quality  multimedia messaging & \multirow {2}{*} {2 Mbps} & \multirow {2}{*} {Mouse} \\
\hline
\end{tabular}
\end{adjustbox}
\end{table}

\subsubsection{User mobility}

Accelerometer data traces in the LiveLab dataset correspond to a 15-minute sampling period. 
We transformed accelerometer data into locations by triangulating expected ranges between the terminals and the access points. Then, since the 15-minute interval was too wide,  we interpolated the locations to simulate sampling intervals of seconds. The conversion of accelerometer data to positions is described in \cite{Mhiri}.

The Position Predictor manages to estimate the position of the terminals with a \SI{93}{\percent} accuracy for a sampling period of 1 second. This accuracy was evaluated by comparing available APs as seen from the estimated position of the terminal with available APs from its actual position (if the sets coincided we considered the estimation to be correct). Even if we increment the sampling period with the goal of reducing control traffic, the Position Predictor achieves an accuracy of an \SI{80}{\percent} for a sampling period of 10 seconds, and a \SI{67}{\percent} accuracy for a sampling period of 30 seconds.

\begin{figure}[!h]
\centering
\includegraphics[width=0.55\textwidth]{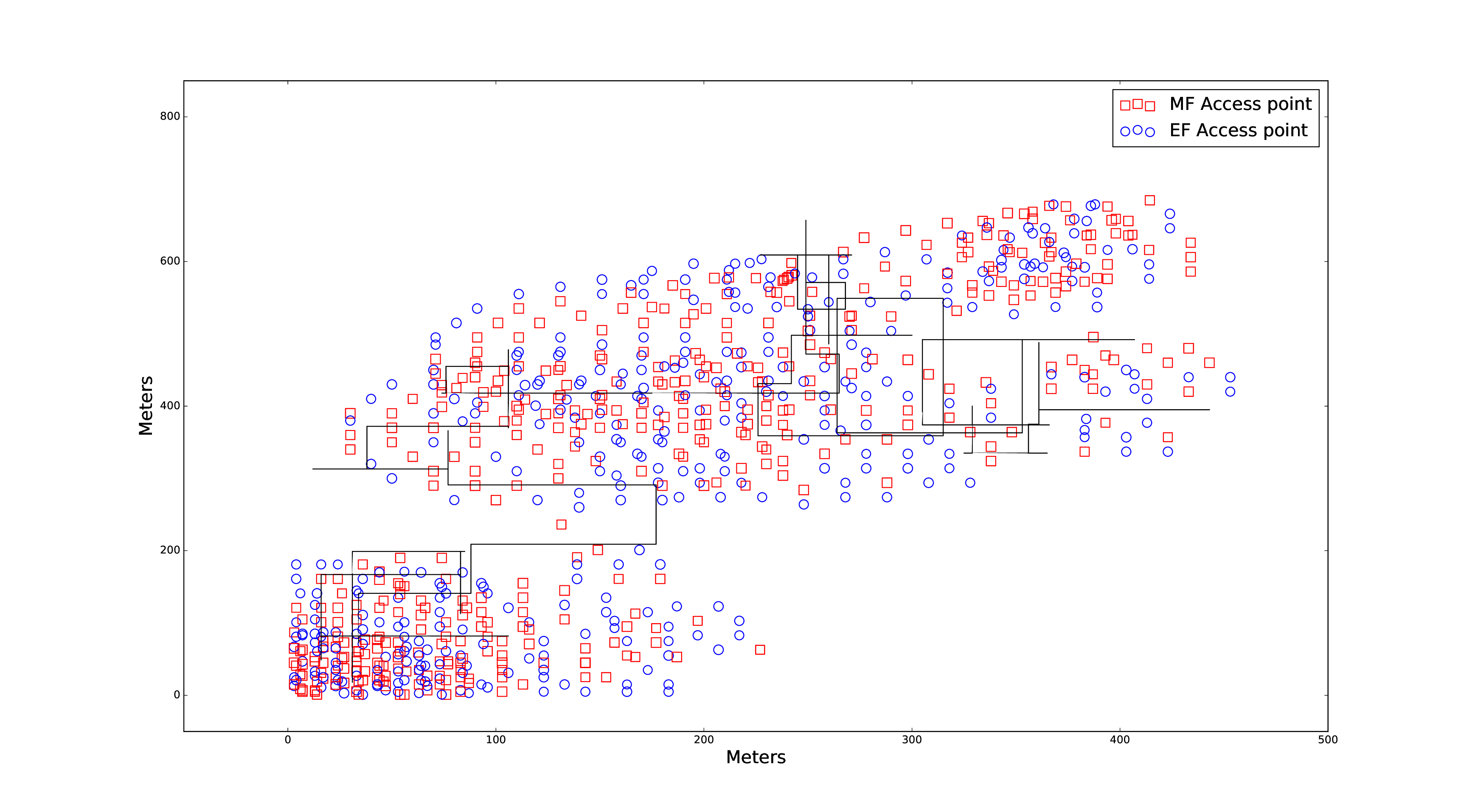}
\caption{Example of estimated user movements across the network}
\label{Saber7}
\end{figure}

Fig. \ref{Saber7} shows the movement of a single user during the 5 working hours in a day. Note the linear patterns of corridors and paths. 

However, we finally assumed that the terminals would not publish their positions in the simulation scenarios, and therefore we estimate $\Omega_i$ from the latest link quality measurements stored in the PHD.

\subsubsection{Prediction of flow type}

To predict flow types at the network side we employed the profiler in \cite{Mhiri} as described in Section \ref{ss_flowPredictor}. Our profiler uses a weight-directed graph to predict the next flow type that a terminal will generate based on its history of application usage. 

The algorithm is able to predict the next flow type a terminal will generate in the scenarios in this section with an average $F$ value of 0.97.

Once the flow type is predicted, the requested rate is taken from the FHD, which keeps average application flow rates.

\subsection{Evaluation Results}
\label{simresults}

To evaluate our AP assignment algorithm  in the scenarios in Section \ref{s_scenario} we compared the following approaches:

\begin{itemize}
\item Centralized optimization with full-knowledge: We assume that a centralized SDN controller optimizes problem (\ref{eq:fittness_funtion})-(\ref{eq:8}) from real-time information of flow types.  
This provides an upper performance bound for both our results \add{and those achievable by \cite{Raschella} (which would rely on more accurate, yet slower to obtain, predictions of flow types)}.   
\item Our first centralized approach based on predicted data: A single SDN controller  for the whole network applies our AP assignment approach from fast predictions of flow types based on historic data and interpolations of terminals' positions.  
\item Our second distributed approach based on predicted data: In this strategy there is one SDN controller per APs cluster. Each controller applies our AP assignment approach inside its cluster. 
\item The terminal-side decision algorithm in \cite{Sood}. A controller sends the spare capacities of the APs to the terminals. The terminals choose the AP within range with highest spare capacity. 
\item Closest-AP scheme, to provide a lowest performance bound.

\end{itemize}

\subsubsection{Algorithm tuning}
\label{adjusAP}

 Instead of choosing a value for $\epsilon$, which could lead to excessively long executions of the assignment algorithm, we studied the evolution of the fitness function as the number of loops/iterations of our algorithm increase, with the aim of establishing a minimum number of iterations that provide a good result. For this purpose, we solved 12000 instances of problem \eqref{eq:fittness_funtion}-\eqref{eq:8}. We increased the number of iterations gradually and studied the fitness achieved. We compared two initialization methods. The first method begins from an empty initial assignment (i.e., $L$ is initially set to {\bf 0}), where no terminal is assigned to an AP. The second method  starts from the previous AP assignment, as long as it is still feasible (i.e, at event k, $L[k]$ is initially set to $L^{*}[k-1]$). In case some connection is lo longer feasible, the corresponding terminal is initially connected to the closest access point.

\begin{figure}[ht!]
\centering
\includegraphics[width=0.5\textwidth]{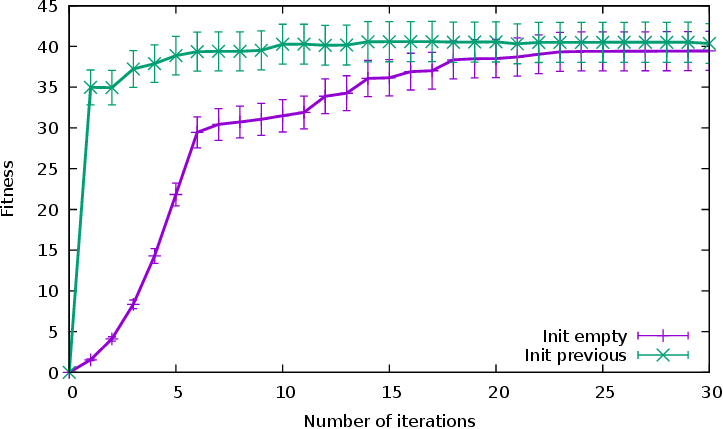}
\caption{Evolution of $f(\cdot)$ with the number of iterations of the algorithm (LiveLab scenario)}
\label{fig:iterations_f_livelab}
\end{figure}

\begin{figure}[ht!]
\centering
\includegraphics[width=0.5\textwidth]{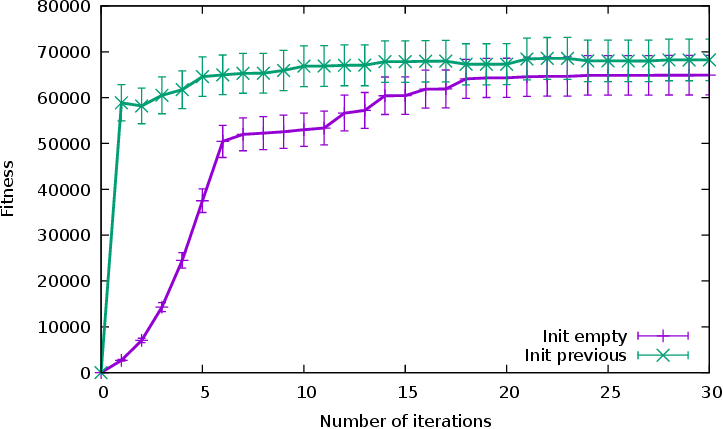}
\caption{Evolution of $f(\cdot)$ with the number of iterations of the algorithm (speculative 5G scenario)}
\label{fig:iterations_f_5g}
\end{figure}

Figs.~\ref{fig:iterations_f_livelab} and~\ref{fig:iterations_f_5g} show the evolution of the fitness function $f(\cdot)$ with the number of iterations of the algorithm in the LiveLab and speculative 5G scenarios, respectively. If the initial assignment was left empty,  the fitness function stabilized  after 20 iterations in both cases. By starting from the previous assignment, however, fitness improved considerably after  5 iterations.

Besides the faster convergence, using the previous assignment for initialization contributes to minimizing the number of reassignments per iteration, hence reducing the control information that must be exchanged through the network and also maximizing assignment stability.

Consequently, in our evaluations in the next sections, at each execution of the algorithm we initialized it with the previous assignment and we ran 5 iterations  in all cases.

\subsubsection{\add{Handover time}}
\label{s_reassingment_latency}

In the worst case, when a terminal launches an application that  the current AP cannot serve, the time to reconfigure the network would be the time to a) detect the new flow ($\chi$), b) notify the controller ($\beta$), c) execute our assignment algorithm ($\gamma$), d) command the terminal to connect to the new AP ($\delta$), and e)  
\add{execute the handover} ($\zeta$). 

As stated before, an SDN architecture is the preferred alternative for our  implementation. In this architecture, terminals and network elements (switches, routers, APs) use \add{SDN southbound protocols} to transmit information on the different traffic flows to the SDN controller, which runs an application that makes decisions. The application also uses \add{SDN protocols} to return management commands \add{to the forwarding devices}.

When a packet arrives to a SDN enabled device it is checked against known flow entries. If the packet does not match any entry, the device sends a copy of the packet to the controller. In \cite{experimentsdn}, the authors analyzed the end-to-end state updating latency of an SDN network with 206 switches and 406 links  in response to an event that caused 1000 flows to be rerouted. In such a scenario, the SDN ONOS controller detected the network event and sent the first Flow-Mod OpenFlow message to reprogram the network in 45.2 ms (median value). A similar experiment was performed in \cite{delaysdn} to study the time required to detect a new flow, send the information to the controller, take a decision,  and insert a new rule in the flow table of the switch/router. In this case, the measured delay for POX and Floodlight SDN controllers was 32 ms and 15 ms, respectively. 

According to \cite{comparacionsdn}, there are other controllers that outperform ONOS, FloodLight and POX in terms of latency and throughput. Therefore, although signaling delay depends on different aspects such as controller implementation and hardware and network load, we can conclude that the time to detect a flow and command a terminal to connect to a particular AP ($\chi + \beta + \delta$) would be in the order of tens of milliseconds.

\begin{figure}[!h]
\centering
\includegraphics[width=0.5\textwidth]{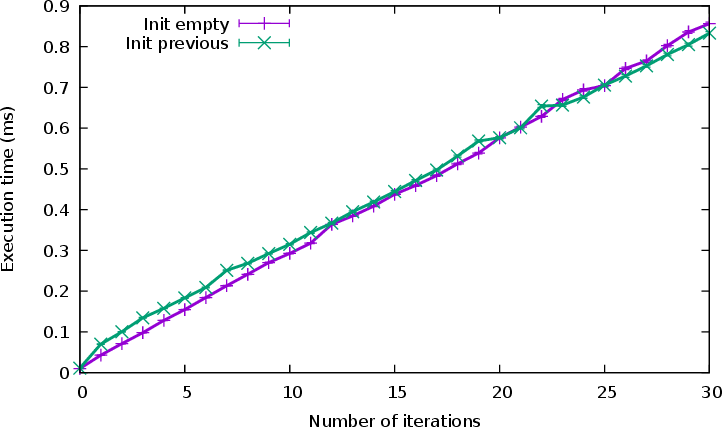}
\caption{Execution time of our assignment algorithm}
\label{fig:execution_time}
\end{figure}

Fig.~\ref{fig:execution_time} represents the time  to execute our assignment algorithm ($\gamma$) on off-the-shelf hardware (Intel Core i7-6700 CPU @ 3.40GHz with 16 GB RAM) in the scenarios in Section \ref{simresults}. The two plots correspond to the two initialization methods in Section \ref{adjusAP}, by averaging 12000 instances of the problem in 10 independent executions.
The first we can notice is that execution time grew linearly with the number of iterations of the algorithm. The only difference was in the initialization stage.
Elapsed time per iteration was less than 0.04 ms. This is because it only depends on the number and types of APs and the number and distribution of users and is independent from traffic demands. Therefore, it was the same in both scenarios in Section \ref{livelabscenario} and  \ref{5Gscenario} for both initialization methods in Section \ref{adjusAP}.

As evaluated in Section \ref{adjusAP}, in the simulation scenarios we just needed five iterations if we restarted the algorithm from the previous-assignment. Thus, 
 $\gamma < 0.2$ ms in practice.

The time to 
\add{execute} the AP handover ($\zeta$) would depend on the particular communication technology used. In a secure Wi-Fi network handover \add{execution} 
may take 6 to 9 s \cite{Wifihandovernormal}, but in an SDN-enabled Wi-Fi network it can be reduced to 34.7 ms on average with some optimizations \cite{WifihandoverSDN}. Average LTE X2-based handover \add{execution} time is about 30 ms  \cite{ltehandovertime}. In 5G networks, this time 
\add{can be reduced to 15.3 ms \cite{latency5g}}.

Summing up, the overall time to optimize the network in the centralized scenarios in Section \ref{simresults} will be dominated by $\chi + \beta + \delta + \zeta$ in the order of tens of miliseconds with Wi-Fi technologies, since $\gamma$ is comparatively negligible.

We remark that there exist mechanisms to reduce or even eliminate handover \add{data} interruption time by keeping the terminals connected all the time at the cost of a higher signaling load). For example, in our previous work in \cite{Giraldo} we maintained  connectivity through 4G cellular communication (which will be likely available in any urban or campus scenario) during Wi-Fi handovers. In \cite{Polese} the authors proposed maintaining connectivity with simultaneous 4G and 5G connections during handovers. A new connection could also be set up with a new AP before the previous connection is released \cite{multinet}.

\subsubsection{Pre-5G LiveLab scenario}
\label{livelabscenario}

Table \add{\ref{tab:table_results} shows the results} of running the algorithms every second over the 34 days (5 hours/day or 18000 samples per day) in the filtered dataset. We compared the traffic losses and \add{average handover times} of the five approaches:

\begin{table}[htb]
\centering
\caption{Simulation results, LiveLab scenario}
\label{tab:table_results}
\begin{tabular}{ | c | c | c | }
\hline
 \textbf{Algorithm} & \textbf{Losses (\si{\percent})} & \textbf{\add{Handover time (ms)}} \\
 \hline
 Centralized (real) & 0.8384 & 80.1 \\
\hline
 Centralized (predicted) & 1.2387 & 80.1 \\
\hline
 Distributed (predicted) & 1.2315 & 80.07 \\
 \hline
 Terminal-side \cite{Sood} & 35.9491 & 34.7 \\
\hline
 Closest-AP & 60.1219 & 34.7 \\

\hline
\end{tabular}
\end{table}

The centralized approach with full knowledge (i.e., using real-time data) achieved the best performance, with 
\SI{0.84}{\percent} traffic losses \add{and a handover time of 80.1 ms}. Our centralized approach with predicted data was able to attain  
\SI{1.24}{\percent} traffic losses \add{and 80.1 ms handover time}. These results for the first two methods were really close, with a difference of just \SI{0.4}{\percent} in losses and \add{for the same handover time}. Thus we verified in the pre-5G  LiveLab scenario that fast flow predictions from historic data are indeed useful for attaining  high performance with low \add{handover} latency. Our distributed approach attained a slightly better performance that our centralized approach (\SI{1.23}{\percent} traffic losses for the same \add{handover time}). This apparently counter-intuitive result \add{is explained by the fact that} in the first iterations, the implicit parallelism of the distributed approach out-weights its more limited exploration capabilities. In other words, it is equivalent to running 15 iterations of the centralized problem instead of 5.

The \add{network} performance of terminal-side decisions with network-side information \cite{Sood} was unsurprisingly lower, with \SI{36}{\percent} traffic losses \add{for 34.7 ms handover time}. This was because  terminals that were close together tended to compete for the same ``best'' APs in a very congested scenario. The results could be improved if the terminals tried different APs until they found one providing satisfactory performance, but they should inform the controller about that, delaying subsequent updates of network-side information. Besides, the time to establish a new connection would be multiplied by the number of retries, affecting \add{handover time, } 
which would be unacceptable to us.

Finally, the lowest bound given by the closest-AP scheme yields
the fact all terminals are allocated to APs in the closest-AP scheme, 
traffic losses over \SI{60}{\percent} and an average \add{handover time of 34.7 ms}. This confirms that there is ample margin for optimization and justifies the interest of applying intelligent decision approaches,  especially in  case of stringent reassignment latency constraints.

\add{Note that average handover times are identical for the centralized approaches, since the only difference among them is that the \textit{real} version uses full-knowledge information about application rates and positions whereas the \textit{predicted} version uses predictions. The distributed algorithm attains an slightly lower handover time due to its distributed operation. On the other hand, both the terminal-side approach and the closest-AP yield lower  
handover times than the network-side approaches. However, despite terminal-side approaches provide a lower handover time it is important to note their packet losses are too high in capacity-demanding  scenarios.}

\subsubsection{Speculative 5G scenario}
\label{5Gscenario}

Table \ref{tab:table_results_5g} shows the results of running the algorithms every second over the 34 days (5 hours/day or 18000 samples per day) in the speculative 5G scenario.

\begin{table}[htb]
\centering
\caption{Simulation results, speculative 5G scenario}
\label{tab:table_results_5g}
\begin{tabular}{ | c | c | c | }
\hline
 \textbf{Algorithm} & \textbf{Losses (\si{\percent})} & \textbf{\add{Handover time (ms)}} \\
\hline
 Centralized (real) & 0.5628 & 60.7 \\
\hline
 Centralized (predicted) & 0.9458 & 60.7 \\
\hline
 Distributed (predicted) & 0.9456 & 60.68 \\
\hline
 Terminal-side \cite{Sood} & 33.5292 & 15.3 \\
\hline
 Closest-AP & 62.4485 & 15.3 \\
\hline

\end{tabular}

\end{table}

The centralized approach with full knowledge achieved again the best performance, with 

\SI{0.56}{\percent} traffic losses and an \add{average handover time of 60.7 ms}, followed by our approaches with predicted data, which attained 

\SI{0.95}{\percent} traffic losses  and  \add{handover time of 60.7 ms}. The difference in this case was \SI{0.4}{\percent} \add{in loss} for the same \add{handover time}.

 The much lower performance of terminal-side decisions with network-side information was \SI{34}{\percent} traffic losses \add{for 15.3 ms of handover time}, for the same reasons as in the pre-5G scenario.
 
 The lowest bound by the closest-AP scheme was \SI{62}{\percent} of packet losses \add{also for a handover time of 15.3 ms}.

 \add{Analogously to the Pre-5G LiveLab scenario, note the difference in handover times between our approaches and terminal-side and closest-AP. Again, the losses out-weight the benefit in reduced latency. Given the fact that the decision time of our algorithms is negligible compared to the other components of handover time, the differences between handover times in the 5G scenario and the LiveLab scenario are explained by the different radio access technologies and the different handover procedures.}

\section{Conclusions}\label{s_conclusion} 
Next-generation ultra-dense wireless networks will have abundant bandwidth resources for current applications, but they will impose \add{more frequent} handovers on user terminals and new applications may push their limits.

Low \add{handover} latency will be a key performance requirement for these new  \add{scenarios. For example, in 5G networks, handovers require the gNBs involved in them to exchange  control messages and user data forwarding may be blocked, so it is essential to optimize handovers by reducing signaling load and time (as they may even cause the interruption of the data flow and decrease the  throughput  of  mobile  users). Therefore} in this paper we have proposed a new AP assignment approach that takes fast decisions based exclusively on the past history of user behavior. It relies on extremely simple and fast methods to predict application flow types from past history, and it is suitable to be implemented with SDN technology. We have evaluated centralized and distributed versions of our approach  with real and speculative 5G data and achieved satisfactory results compared to  centralized optimization with full real-time knowledge of the network and terminal-side decisions based on network-side information.  

\section*{Acknowledgments}

This research has been partially supported with MINECO grants TEC2016-76465-C2-2-R and RTC-2016-4898-7, Xunta de Galicia grant GRC2018/53 and ``la Caixa" Foundation (ID 100010434) fellowship code LCF/BQ/ES18/11670020, Spain.

\ifCLASSOPTIONcaptionsoff
  \newpage
\fi


\begin{IEEEbiography} [{\includegraphics[width=1in,height=1.25in,clip, keepaspectratio]{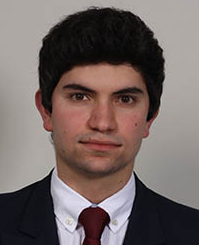}}]{Pablo Fondo-Ferreiro}
received his Bachelor's Degree in Telecommunication Engineering in 2016 from the University of Vigo, receiving the award for the best academic record. In 2018 received his Master's Degree in Telecommunication Engineering from the same university, holding the best academic record. In 2016 he received a collaboration grant from the Spanish Ministry of Education for research on Software Defined Networking and Energy Efficiency in Communication Networks, whose results have been submitted for publication. In 2018 he received a fellowship from "la Caixa" Foundation to pursue his PhD in Information and Communication Technologies at University of Vigo. His research interests include SDN, Mobile Networks and Artificial Intelligence.
\end{IEEEbiography}
\begin{IEEEbiography} [{\includegraphics[width=1in,height=1.25in,clip, keepaspectratio]{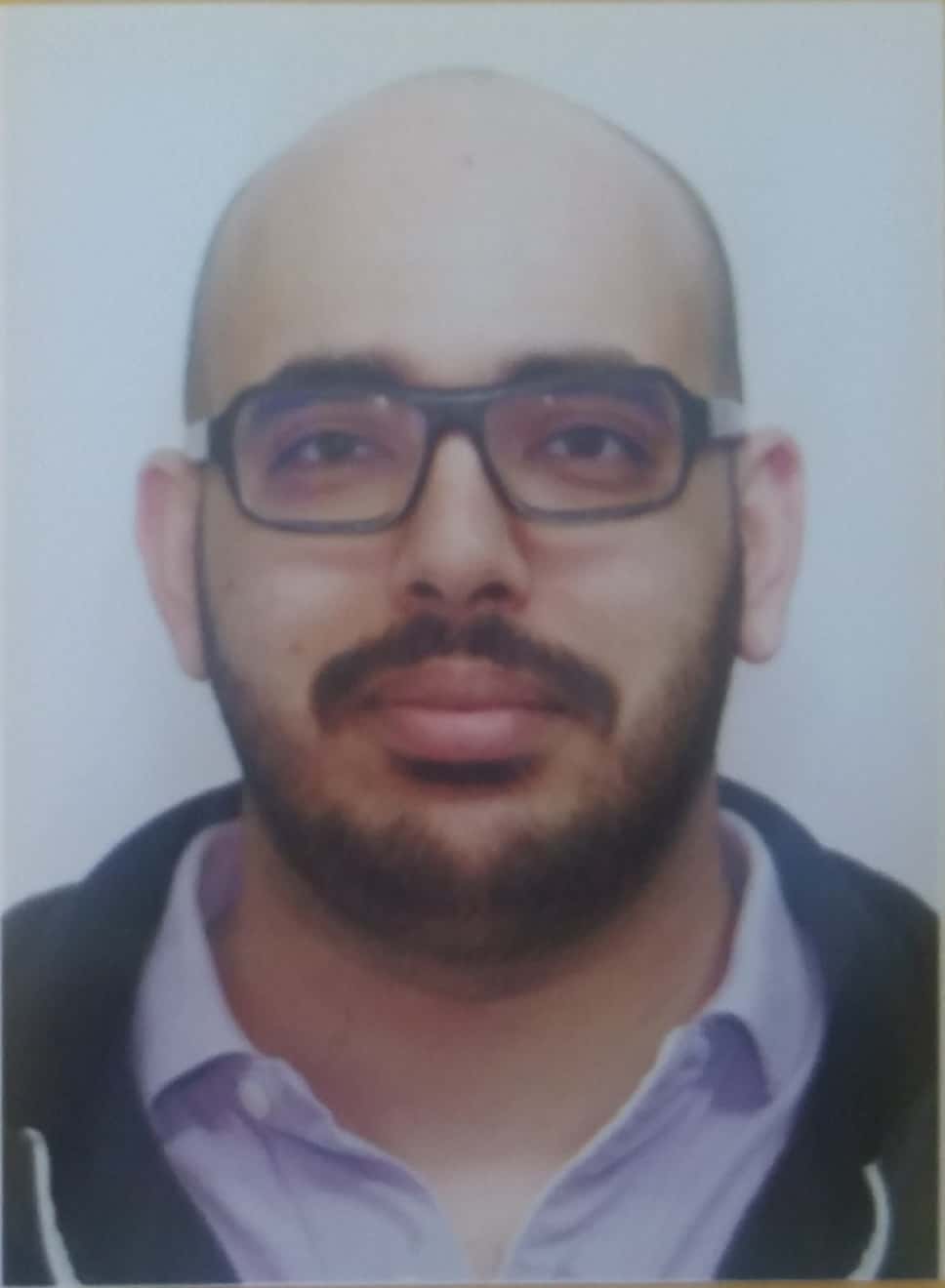}}]{Saber Mhiri}
received his research Master degree in computer science and multimedia in 2014. Since then he is a PhD student working with the GTI Group,  University of Vigo, Spain. He participates in the international Erasmus program EGOV-TN funded by the European  Union. He has authored or coauthored of several papers in international conferences in the fields of networks, distributed systems, telecommunications and computer science. 
\end{IEEEbiography}

\begin{IEEEbiography} [{\includegraphics[width=1in,height=1.25in,clip, keepaspectratio]{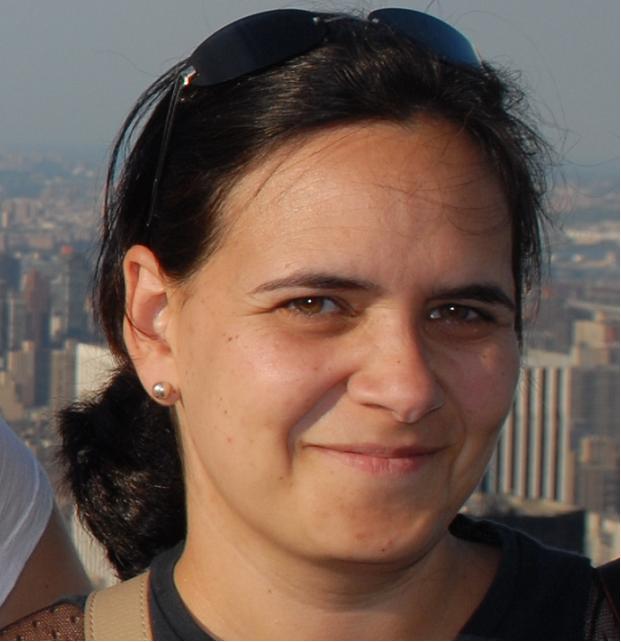}}]{Cristina L\'opez Bravo}
received her Telecommunications Engineering and Ph.D. degrees in 2000 and 2004, respectively. She has been an associate professor with the Departamento de Enxe\~nar\'ia Telem\'atica, Universidade de Vigo, since 2008. She has authored or coauthored over 30 papers in international refereed journals and international conferences, in the fields of telecommunications and computer science. She has participated in several relevant European (EphotonOne, Bone) and  national projects (ARPAq, CAPITAL, CALM, COINS, AIMS).
\end{IEEEbiography}


\begin{IEEEbiography}[{\includegraphics[width=1in,height=1.25in,clip,keepaspectratio]{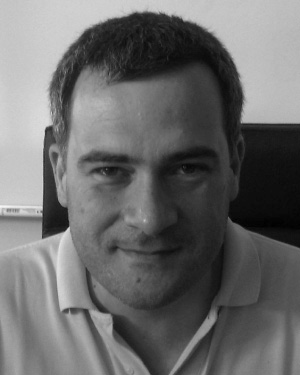}}]{Francisco Javier Gonzal\'ez Casta\~no}
is currently a Catedr\'atico de Universidad (Full Professor) with the Telematics Engineering Department, University of Vigo, Spain, where he leads the Information Technology Group. He has authored over 80 papers in international journals, in the fields of telecommunications and computer science, and has participated in several relevant national and international projects. He holds two U. S. patents.
\end{IEEEbiography}

\begin{IEEEbiography}[{\includegraphics[width=1in,height=1.25in,clip,keepaspectratio]{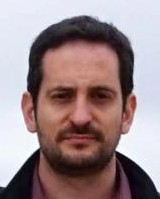}}]{Felipe Gil-Castiñeira}
Felipe Gil-Castiñeira received the M.Sc. degree in telecommunication engineering (major in telematics)
and the Ph.D. degree in telecommunication engineering from University of Vigo in 2002 and 2007, respectively. He is
currently an Associate Professor with the Department of Telematics Engineering, UVigo. Between 2014
and September 2016 he was the head of the Intelligent Networked Systems area in Gradiant (Galician
Research and Development Center in Advanced Telecomunications). His research interests include
wireless communication technologies and core network technologies, multimedia communications,
embedded systems, ubiquitous computing, and the Internet of things. He has published over  sixty
papers in these fields in international journals and conference proceedings. He has led several national and
international R\&D projects. He holds two patents related to mobile communications. He is the co-founder of a
University spin-off, and has recently been awarded with funding from the regional government to transfer
ubiquitous computing technology developed at the University of Vigo to the market.
\end{IEEEbiography}




\end{document}